\documentclass{PoS}

\title{Nucleon isovector charges from physical mass domain-wall QCD}

\ShortTitle{DWF nucleon charges}

\author{\speaker{Shigemi Ohta}for the LHP, RBC, and UKQCD Collaborations\\
        Institute of Particle and Nuclear Studies, KEK, Tsukuba, Ibaraki, 305-0801, Japan\\
        Department of Particle and Nuclear Studies, SOKENDAI, Hayama, Kanagawa, 240-0193, Japan\\
        RIKEN-BNL Research Center, BNL, Upton, NY, 11973, USA\\
        E-mail: \email{shigemi.ohta@kek.jp}}

\abstract{
\vspace{-129mm}\parbox{\textwidth}{\flushright\large\rm \hfill KEK-TH-2161, RBRC-1319}\vspace{122mm}
Systematics in nucleon isovector vector, \(g_V\), and axialvector, \(g_A\), charges calculated on a 2+1-flavor dynamical domain-wall-fermions (DWF) ensemble at physical mass jointly generated by RIKEN-BNL-Columbia (RBC) and UKQCD Collaborations with lattice cut off of 1.730(4) GeV, are analyzed.
Both are calculated with about a percent or less statistical errors.
A few standard-deviation systematics seen in vector charge is consistent with possible \(O(a^2)\) discretization error through small excited-state contamination.
Axialvector charge is found with three to nine standard-deviation systematic deficit, compared with experiments, depending on calculation methods.
}

\FullConference{37th International Symposium on Lattice Field Theory - Lattice2019\\
		16-22 June 2019\\
		Wuhan, China}

\begin{document}

\section{Introduction}

Domain-wall fermions (DWF) maintain continuum-like flavor and chiral symmetries on the lattice up to \(O(a^2)\) discretization effects in lattice spacing \(a\), because \(O(a)\) effect is exponentially suppressed in the domain-wall fifth dimension.
The RBC and UKQCD Collaborations have been jointly generating 2+1-flavor dynamical DWF lattice-QCD ensembles for a time \cite{Blum:2000kn,Aoki:2004ht,Allton:2008pn,Aoki:2010dy,Arthur:2012yc,Blum:2014tka}.
The joint Collaborations reached the physical mass \cite{Blum:2014tka} several years ago, and since then have been producing interesting results on pion and kaon physics and muon anomalous magnetic moment.
The joint Collaborations also have been working on nucleon structure using these 2+1-flavor dynamical DWF ensembles \cite{Yamazaki:2008py,Yamazaki:2009zq,Aoki:2010xg,Lin:2014saa,Ohta:2013qda,Ohta:2014rfa,Ohta:2015aos,Abramczyk:2016ziv,Ohta:2017gzg}.
More recently the LHP Collaboration started working with the joint Collaborations in calculating nucleon structure using the physical-mass DWF ensembles \cite{Syritsyn:2014xwa}.
Last year at this conference in East Lansing I reported  \cite{Ohta:2018zfp} the status of nucleon isovector vector, \(g_V\),  and axialvector, \(g_A\), charges calculated on a physical-mass 2+1-flavor dynamical DWF ensemble at lattice cut off of about 1.730(4) GeV  \cite{Blum:2014tka} along with isovector scalar, \(g_S\), and tensor, \(g_T\), couplings.
This year I like to add some analyses on systematics seen in the vector and axialvector charges.

Standard DWF local-current definitions are used for the quark isovector currents.
Necessary non-perturbative renormalizations for these local-current bilinears have been worked out in the meson sector \cite{Blum:2014tka}.
The present result for the vector-charge renormalization provides a further opportunity to scrutinize this renormalization.
In contrast, the calculation of the axialvector charge, now experimentally known as \(g_A/g_V = 1.2732(23)\)  \cite{PhysRevD.98.030001} or 1.2764(6) \cite{Markisch:2018ndu} has been more problematic:
The RBC Collaboration have been reporting  deficit in \(g_A\) \cite{Yamazaki:2008py,Yamazaki:2009zq,Aoki:2010xg,Lin:2014saa}, and values such as about 1.15(5) were reported from ensembles with heavier than physical up- and down-quark mass \cite{Ohta:2013qda,Ohta:2014rfa,Ohta:2015aos,Abramczyk:2016ziv,Ohta:2017gzg}.
These observations appear to have been confirmed by several other major collaborations \cite{Dragos:2016rtx,Bhattacharya:2016zcn,Liang:2016fgy,Ishikawa:2018rew,Chang:2018uxx} using different actions but with similar lattice spacings and quark masses, though extrapolations to physical mass seem to differ.
Especially important for calculations with Wilson-fermion quarks \cite{Dragos:2016rtx,Bhattacharya:2016zcn,Ishikawa:2018rew} was to remove the \(O(a)\) discretization systematic errors \cite{Liang:2016fgy}. 
Possible causes of this systematics have been extensively discussed, such as excited-state  contaminations and finite lattice volume.
The RBC Collaboration were yet to see any evidence that their vector and axialvector charges suffer any excited-state contamination.
Note also that it is well known from the days of the MIT bag model \cite{Chodos:1974pn} that the so-called ``pion cloud'' around nucleon would be important for this observable, and proper account of its geometry \cite{Adkins:1983ya}.

We use the ``48I'' \(48^3\times 96\) 2+1-flavor dynamical M\"{o}bius DWF ensemble at physical mass with Iwasaki gauge action of \(\beta=2.13\), or of lattice cut off of  \(a^{-1} = 1.730(4)\) GeV, jointly generated by the RBC and UKQCD Collaborations \cite{Blum:2014tka}.
In total 130 configurations, separated by 20 MD trajectories in the range of trajectory number 620 to 980 and by 10 MD trajectories in the range of trajectory number from 990 to 2160, except the missing 1050, 1070, 1150, 1170, 1250, 1270, and 1470, are used.
Each configuration is deflated with 2000 low Dirac eigenvalues \cite{Clark:2017wom}.
The ``AMA'' statistics trick  \cite{Shintani:2014vja}, with \(4^4=256\) AMA sloppy samples unbiased by 4 in time precision ones from each configuration, is used.
Gauge-invariant Gaussian smearing  \cite{Alexandrou:1992ti,Berruto:2005hg} with similar parameters as in the past RBC nucleon structure calculations is applied to nucleon source and sink, separated by \(8 \le T \le 12\) in time.

\section{Vector charge}

\begin{figure}[tb]
\begin{center}
\includegraphics[width=.49\textwidth,clip]{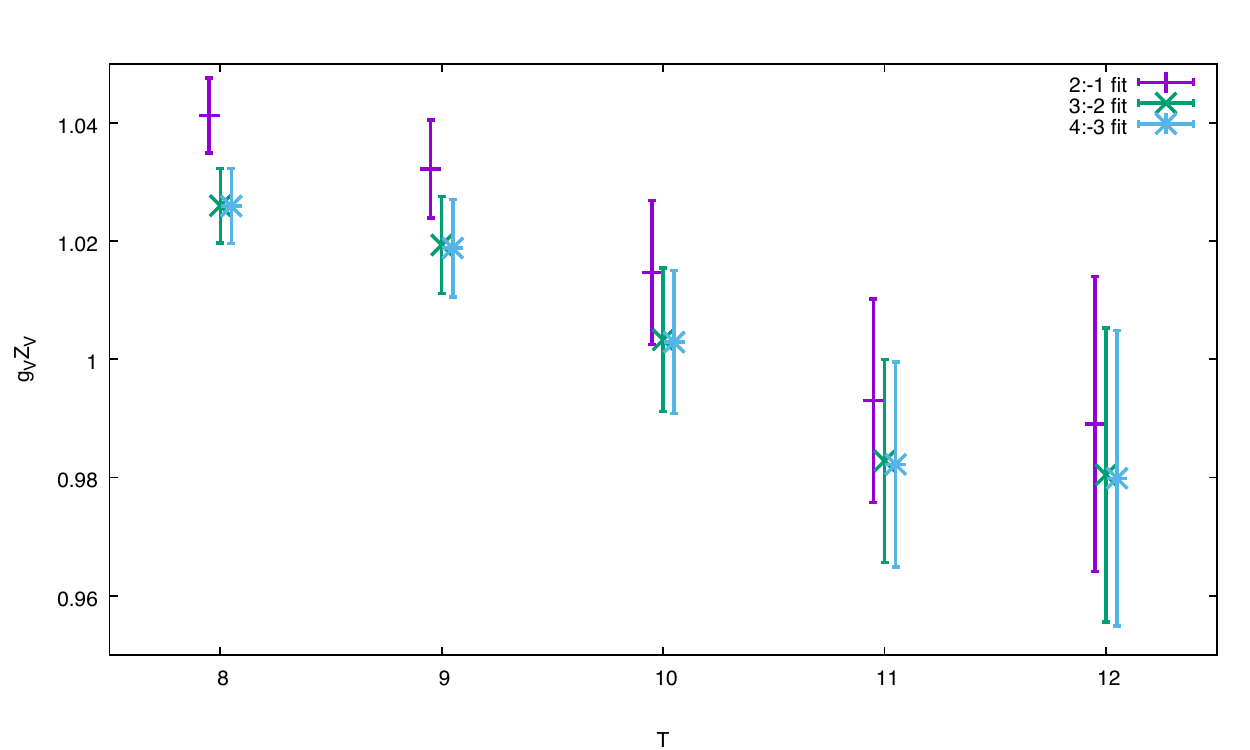}
\includegraphics[width=.49\textwidth,clip]{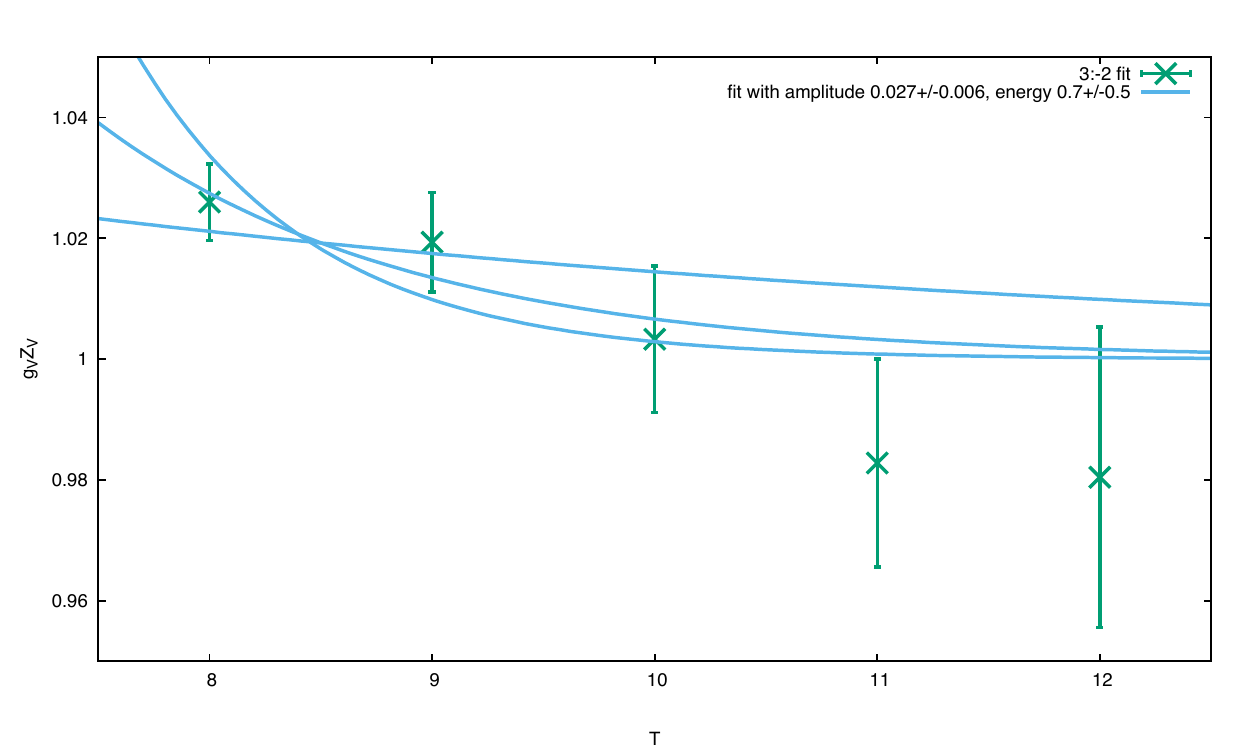}
\end{center}
\caption{\label{fig:gV}
Nucleon isovector vector charge, \(g_V\), with renormalization, \(Z_V = 0.71076(25)\) from meson sector.
Left: fits to the plateaux with 2, 3, and 4 points closest to source or sink eliminated, vs.\  source-sink separation, \(T\).
Right: possible single-excited-state fits to the values for first and last 3-points removal.
}
\end{figure}

The results for the isovector vector charge, \(g_V\), obtained from the local current, are presented in Fig.\ \ref{fig:gV},
with renormalization, \(Z_V = 0.71076(25)\) obtained in the meson sector \cite{Blum:2014tka}.
On the left are constant fits with elimination of first and last two, three, and four points close to the source or sink plotted against the source-sink separation of \(T=8\), 9, 10, 11, and 12.
The statistical errors of these plateau fits are about a percent or less.

As can be seen the results with first and last two-point elimination seem to systematically deviate from fits with three- and four-point elimination, while the latter two agree well with each other.
Though not quite statisitically significant, this deviation likely signals leak through the DWF fifth dimension.
The results from the source-sink separations of \(T=8\) and 9 deviate from unity by a few percent, or equivalently by a few standard deviations, while the results from the longer separations capture unity within their respective statistical errors.

These deviations can be attributed to excited-state contamination through discretization that is expected at \(O(a^2)\), namely of a couple of percent.
For this to happen there must be some excited states present in the smeared nucleon source or sink with finite amplitude.
A higher-dimensional \(O(a^2)\) discretization term in the local current that is not diagonal between the ground and excited states picks up such mixing of excited states to result in what we see as the deviation from unity here.
If confirmed, this would be the first time we see such contamination in this quantity, the charge of a conserved current.
Thus we attempted to fit the deviations to a single-excited-state form of \(A \exp(-B(T-8))\) and obtained \(A=0.027(6)\) and \(B=0.7(5)\).
The latter is to be compared with supposedly near-by excited states of a single pion with one momentum unit, \(m_\pi a + 2\pi/L = 0.08+0.13 = 0.21\), or two pions with zero momentum, \(2 m_\pi a = 0.16\).
The possibility for those excited states contaminating is not excluded.
However seemingly steeper slope at the separation \(T\) of 9--10 than of 8--9 prevents us from firmly concluding, as a single excited state that is more rapidly decaying than the ground state would result in a shallower slope at larger separation.

A possible explanation for the steeper slope at larger separation is that the nucleon ground-state signal itself is beginning to erode there, by \(T=10\): larger statistical errors here and also in the nucleon effective mass \cite{Ohta:2018zfp} in this time range suggest this as the likeliest cause.
Calculations at shorter source-sink separations such as \(T=7\) and 6 would be useful, so are planned in the nearest future.
The separations shorter than 6 are impractical given the amount of fifth-dimensional leak.
Until such shorter-separation calculations are performed, the data from the larger separations of \(T=10\), 11, and 12 will stay rather useless.

\section{Axialvector charge}

\begin{figure}[t]
\begin{center}
\includegraphics[width=.49\textwidth,clip]{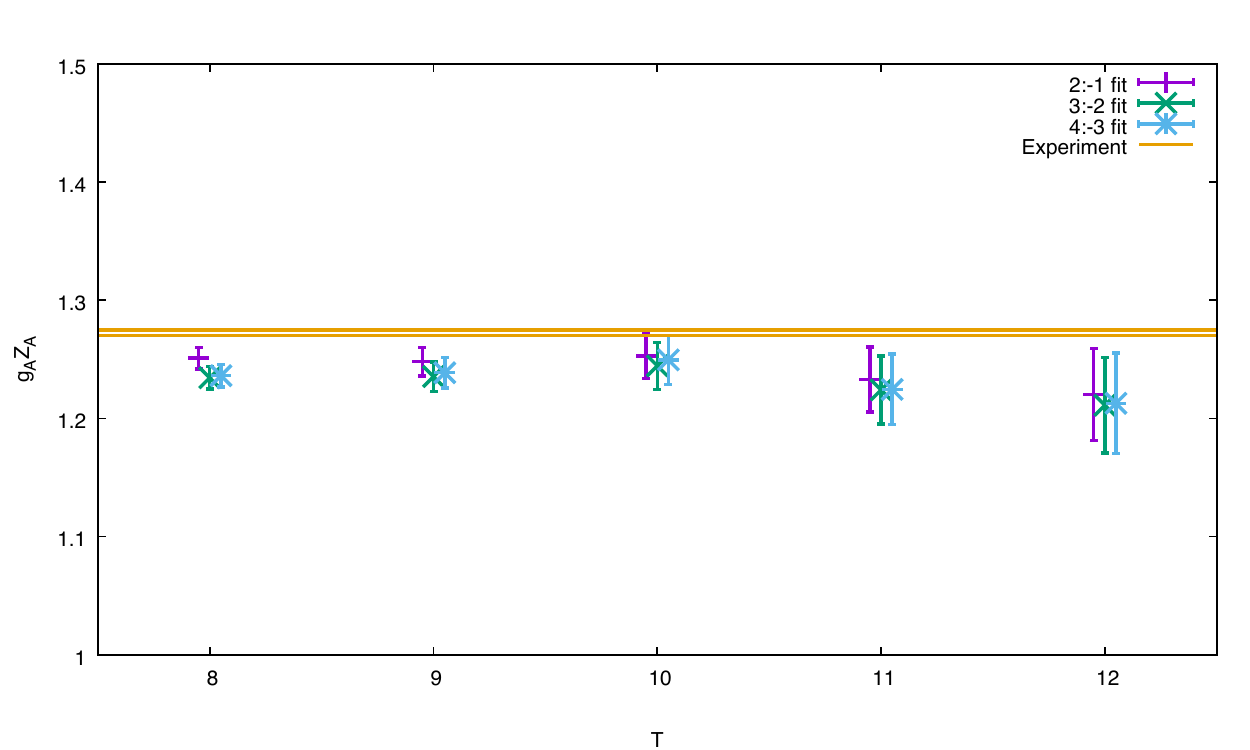}
\includegraphics[width=.49\textwidth,clip]{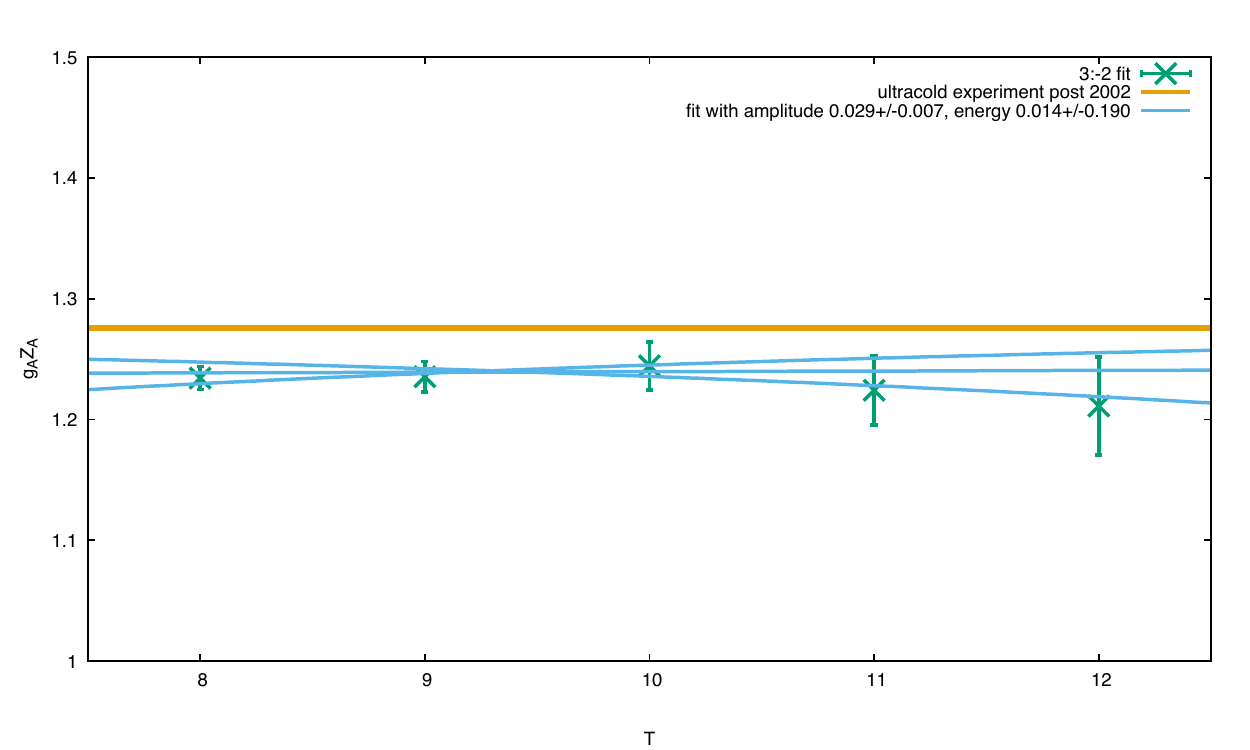}
\end{center}
\caption{\label{fig:gA}
Nucleon isovector axialvector charge, \(g_A\), with renormalization, \(Z_A = 0.71191(5)\) from meson sector.
Left: fits to the plateaux with 2, 3, and 4 points closest to source or sink eliminated, vs.\  source-sink separation, \(T\).
Right: possible single-excited-state fits to the values for first and last 3-points removal.
}
\end{figure}

The results for the isovector axialvector charge, \(g_A\), obtained from the local current, are presented in Fig.\ \ref{fig:gA}, with renormalization, \(Z_A = 0.71191(5)\), obtained in the meson sector \cite{Blum:2014tka}.
Values obtained by constant fit eliminating first and last two, three, and four points close to the source or sink are plotted against the source-sink separation, \(T\).
Here again a slight systematic difference between plateau fits with eliminations of first and last two points from the ones with first and last three or four points is seen.

In contrast to the vector charge, here with the axialvector charge there is no appreciative dependence on the source-sink separation:
a few-standard-deviation deficit in comparison with the experiment appears solid.
In other words we do not see any evidence for excited-state contamination here.
Attempted single-excited state fit resulted in an excitation-energy estimate of 0.014(19) consistent with zero.

\begin{figure}[t]
\begin{center}
\includegraphics[width=.49\textwidth,clip]{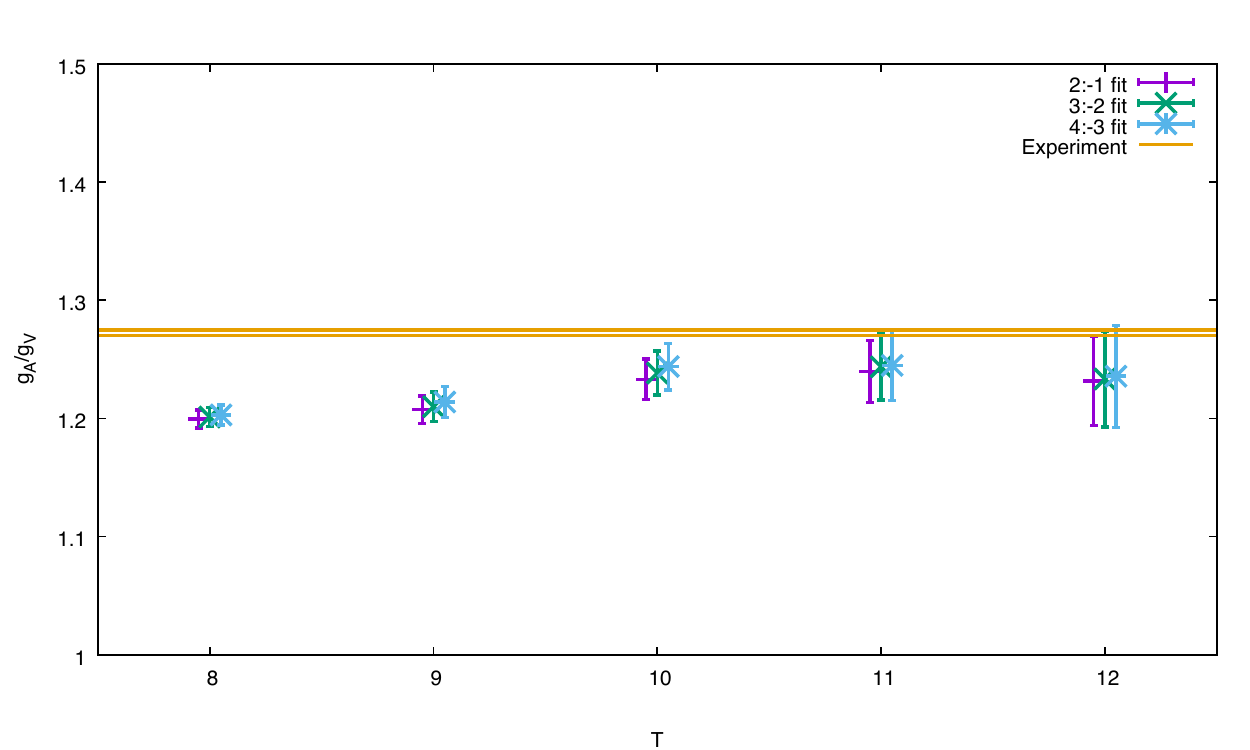}
\includegraphics[width=.49\textwidth,clip]{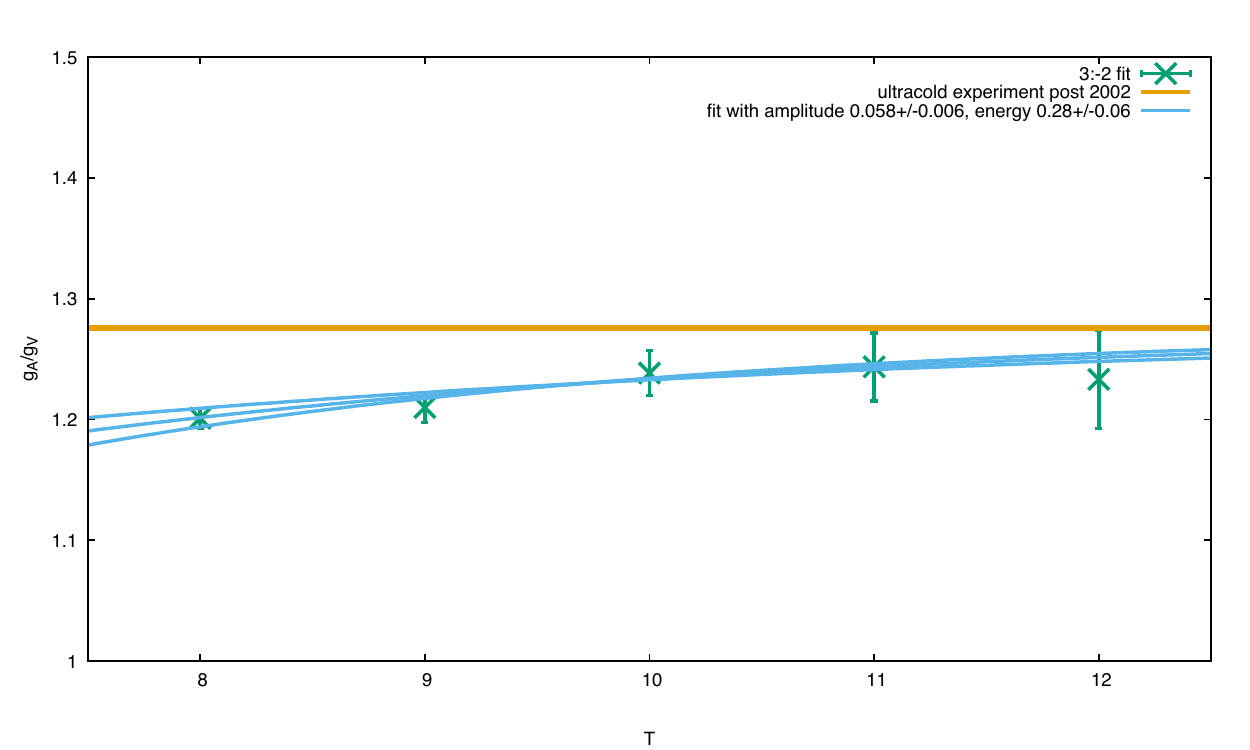}
\end{center}
\caption{\label{fig:AV}
Ratio, \(g_A/g_V\), of nucleon isovector axialvector to vector charges from direct ratio of respective three-point functions and involving meson-sector renormalizations.
Left: fits to the plateaux with 2, 3, and 4 points closest to source or sink eliminated, vs.\  source-sink separation, \(T\).
Right: possible single-excited-state fits to the values for first and last 3-points removal.
}
\end{figure}

Because DWF preserve the chiral symmetry, local vector and axialvector currents share common renormalization, \(Z_V=Z_A\),  up to \(O(a^2)\) discretization effects.
Indeed in meson sector \(Z_V = 71076(25)\) and \(Z_A = 0.71191(5)\) were obtained \cite{Blum:2014tka}.
Thus we can use a ratio of the local axialvector- and vector-current three-point functions for estimating the axialvector charge without involving meson-sector renormalizations.
The results for this alternative estimation are shown in Fig.\ \ref{fig:AV}.
Again values obtained by constant fit eliminating first and last two, three, and four points close to the source or sink are plotted against the source-sink separation, \(T\).
With this method the systematics arising from plateau selection or DWF fifth-dimensional leak disappears reflecting good chiral symmetry of the underlying ensemble.
Here the deficit from the experimental value at shorter separation grows to several standard deviation.
Again the slope from T=9 to 10 is steeper than that from 8 to 9.
Yet a single-excited-state fit results in excitation estimate of 0.28(6).

Both methods result in significant deficit in comparison with experiments, especially at shorter source-sink separations of \(T=8\) and 9 where also the two methods disagree with each other (see Table \ref{tab:gA}).
\begin{table}[b]
\begin{center}
\begin{tabular}{lll}
\hline\hline
\multicolumn{1}{c}{T}&
\multicolumn{1}{c} {\(Z_Ag_A\)}&
\multicolumn{1}{c}{\(g_A/g_V\)}\\
\hline
8 & 1.234(9) & 1.201(8)\\
9 & 1.235(13)& 1.210(12)\\
\hline\hline
\end{tabular}
\caption{
\label{tab:gA}
Isovector axialvector charge with meson-sector renormalization \(Z_A=0.71191(5)\) and as a direct ratio to the vector charge for source-sink separation \(T\).
Fits are made eliminating three closest points to source or sink.
All values show significant deficit in comparison with experiments such PDG estimate \cite{PhysRevD.98.030001} of \(g_A/g_V=1.2732(23)\) or the latest UCN \cite{Markisch:2018ndu} of \(1.2764(6)\).
}
\end{center}
\end{table}
For larger separations, \(T \ge 10\), the difference between the two calculational methods, if any, is hidden by larger statistical noises.

Those with meson-sector renormalization are closer to experiments yet at shorter source-sink separation of \(T=8\) the deficit is more than four standard deviations and at longer separation of 9 three standard deviations if compared with conservative PDG estimate \cite{PhysRevD.98.030001} of \(g_A/g_V=1.2732(23)\).
If compared with the latest ultra-cold-neutron experiment \cite{Markisch:2018ndu} of \(g_A/g_V=1.2764(6)\) alone, the deficits grow to five and four standard deviations.
Values from direct ratio to vector charge show larger deficits, more than nine standard deviations at shorter and more than five at longer separations.
Larger deficits here for the shorter separation may be caused by the suspected excited-state contamination in vector charge that is in the denominator.
However such an interpretation leaves a question of how excited states affect vector and axialvector charge so differently.

Pending the planned shorter-separation study that can clear this question, a more conservative estimate of isovector axialvector charge can be obtained from the larger separation of \(T=9\) by averaging the two methods and using one half of the difference as systematic error: \(g_A/g_V = 1.222(13)_{\rm stat}(13)_{\rm syst}.\)
Even with this the deficit is three-standard-deviation significant.

\section{Summary}

Deviation from unity of a few percent seen in renormalized vector charge can be fit by a single-excited-state exponential form, but the obtained excitation estimate of 0.7(5) in lattice unit is too broad to be conclusive.

\begin{figure}[t]
\begin{center}
\includegraphics[width=.49\textwidth,clip]{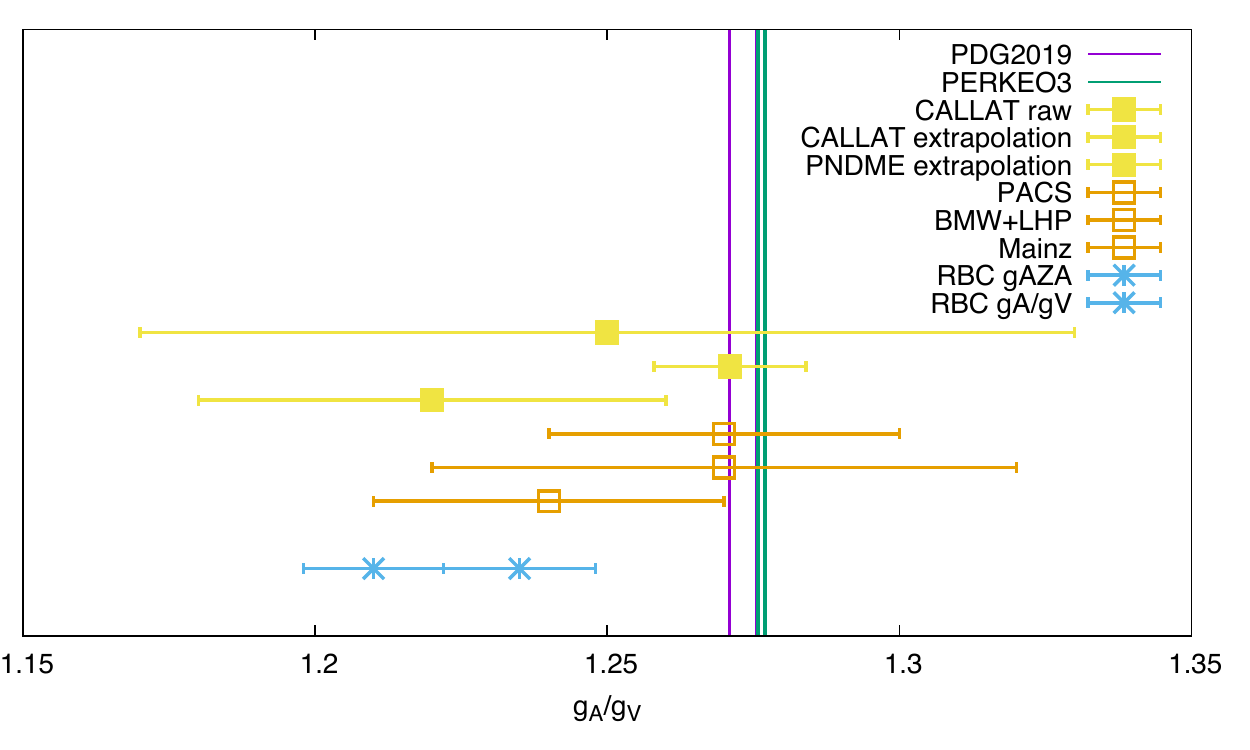}
\caption{
\label{fig:AVcomparison}
Nucleon isovector axialvector charge estimators in this work with source-sink separation of \(T=9\) (\(\times\)) are compared with experiments (vertical solid lines)  \cite{PhysRevD.98.030001,Markisch:2018ndu} and some recent lattice calculations by other collaborations that were available before the conference: nonunitary calculations by CALLAT \cite{Chang:2018uxx}, at physical point and extrapolation from heavy masses, and by PNDME \cite{Bhattacharya:2016zcn}, and unitary ones with various Wilson fermions \cite{Shintani:2018ozy,Hasan:2019noy,Harris:2019bih}
}
\end{center}
\end{figure}

Deficit seen in axialvector charge compared with experiments is significant:
a conservative estimates from source-sink separation of \(T=9\) is three standard-deviation away from experiments, while individual values show five to nine standard-deviation deficit.
It is interesting to compare this conservative estimate with some recent lattice calculations from other collaborations, as in Fig.\ \ref{fig:AVcomparison}.

The author thanks the members of LHP, RBC, and UKQCD Collaborations, and in particular Sergey Syritsyn.
The ``48I'' ensemble was generated using the IBM Blue Gene/Q (BG/Q) ``Mira'' machines at the Argonne Leadership Class Facility (ALCF) provided under the Incite Program of the US DOE, on the ``DiRAC'' BG/Q system funded by the UK STFC in the Advanced Computing Facility at the University of Edinburgh, and on the BG/Q  machines at the Brookhaven National Laboratory.
The nucleon calculations were done using ALCF Mira.
The author was partially supported by Japan Society for the Promotion of Sciences, Kakenhi grant 15K05064.

\bibliographystyle{epj}
\bibliography{nucleon}



\end{document}